\documentclass[12pt,dvipsnames]{article}
\usepackage[top=80pt,bottom=85pt,left=85pt,right=85pt]{geometry}

\pdfoutput=1
\usepackage[utf8]{inputenc}
\usepackage[T1]{fontenc} 
\usepackage[english]{babel} 
\usepackage{braket}
\usepackage[dvipsnames,table]{xcolor} 
\usepackage{physics}
\usepackage{tabularx} 
\usepackage{ragged2e}
\usepackage{rotating}
\usepackage{makecell}
\usepackage{multirow} 
\usepackage{comment} 
\usepackage{amsmath}
\usepackage{bbm}
\usepackage{cite}
\usepackage[all]{xy}

\usepackage{tikz}
\usetikzlibrary{math} 
\usetikzlibrary{fit,backgrounds}
\pgfdeclarelayer{bg}
\pgfsetlayers{bg,main}
\usetikzlibrary{3d} 
\usepackage{tikz-3dplot}
\usepackage{tikz-cd} 
\usepackage{microtype}
\usetikzlibrary{arrows.meta,positioning} 
\usepackage{amsthm}
\usepackage{mathrsfs} 
\usepackage[strict]{changepage}
\newcommand\scalemath[2]{\scalebox{#1}{\mbox{\ensuremath{\displaystyle #2}}}}

\tikzcdset{
  cells={font=\everymath\expandafter{\the\everymath\displaystyle}},
}

\usepackage{amssymb} 
\usepackage{subcaption} 
\DeclareCaptionFormat{custom}

\usepackage[pdftex,colorlinks=true]{hyperref} 
\hypersetup{urlcolor=MidnightBlue, citecolor=red, linkcolor=MidnightBlue}

\usepackage{float} 
\usepackage[textsize=tiny]{todonotes}

\usepackage[capitalise]{cleveref}

\numberwithin{equation}{section}

\definecolor{cambridgeblue}{rgb}{0.64, 0.76, 0.68}
\definecolor{caribbeangreen}{rgb}{0.0, 0.8, 0.6}
\definecolor{celadon}{rgb}{0.67, 0.88, 0.69}
\definecolor{champagne}{rgb}{0.97, 0.91, 0.81}
\definecolor{cream}{rgb}{1.0, 0.99, 0.82}
\definecolor{cyan(process)}{rgb}{0.0, 0.72, 0.92}
\definecolor{brilliantlavender}{rgb}{0.96, 0.73, 1.0}
\definecolor{candypink}{rgb}{0.89, 0.44, 0.48}

\usepackage{pdflscape}
\usepackage{paralist}

\usetikzlibrary{positioning}
\usetikzlibrary{arrows}
\usetikzlibrary{decorations.pathreplacing}
\usetikzlibrary{shapes.geometric}
\usetikzlibrary{calc}
\usetikzlibrary{decorations.pathmorphing}
\usetikzlibrary{shapes.misc}

\tikzset{gaugeSU/.style={inner sep=1.7mm,draw=none,fill=yellow,minimum size=2mm,circle, draw}}
\tikzset{flavourSU/.style={draw=none,minimum size=2mm,fill=white, regular polygon,regular polygon sides=4,draw}}
\tikzset{flavour/.style={draw=none,minimum size=0.3mm,fill=white, regular polygon,regular polygon sides=4,draw}}
\tikzset{gaugeBig/.style={inner sep=1mm,draw=none,fill=white,minimum size=2mm,circle, draw}}
\tikzset{bd/.style={circle, draw=black, inner sep=0pt, fill=black, minimum size=2mm}}
\tikzset{wd/.style={circle, draw=black, inner sep=0pt, fill=white, minimum size=2mm}}
\tikzset{Dynkin/.style={circle, draw=black, inner sep=0pt, fill=white, minimum size=2mm}}
\tikzstyle{ligne}=[draw, very thick] 
\tikzstyle{gridline}=[draw, gray] 
\tikzset{gauge/.style={circle, draw,inner sep=2.5pt,fill=white}}
\tikzset{gaugeo/.style={circle, draw,inner sep=2.5pt,fill=orange}}
\tikzset{gaugec/.style={circle, draw,inner sep=2.5pt,fill=cyan}}
\tikzset{gauger/.style={circle, draw,inner sep=2.5pt,fill=red}}
\tikzset{gaugeb/.style={circle, draw,inner sep=2.5pt,fill=blue}}
\tikzset{gaugeg/.style={circle, draw,inner sep=2.5pt,fill=green}}
\tikzset{gaugem/.style={circle, draw,inner sep=2.5pt,fill=magenta}}
\tikzset{gaugey/.style={circle, draw,inner sep=2.5pt,fill=yellow}}
\tikzset{hasse/.style={circle, fill,inner sep=2pt}}
\tikzset{shrinky/.style={circle, fill,inner sep=1pt}}
\tikzset{sized/.style={circle, draw, inner sep=1.5pt}}
\tikzset{seven/.style={circle, draw,inner sep=3pt}}

\tikzset{dotto/.style={circle, orange, draw,inner sep=1.5pt,fill=orange}}
\tikzset{dottp/.style={circle, purple, draw,inner sep=1.5pt,fill=purple}}
\tikzset{dottc/.style={circle, cyan, draw,inner sep=1.5pt,fill=cyan}}
\tikzset{dottr/.style={circle, red, draw,inner sep=1.5pt,fill=red}}
\tikzset{dottb/.style={circle, blue, draw,inner sep=1.5pt,fill=blue}}
\tikzset{dottg/.style={circle, green, draw,inner sep=1.5pt,fill=green}}
\tikzset{dottm/.style={circle, magenta, draw,inner sep=1.5pt,fill=magenta}}

\usepackage{ifthen}
\usepackage{xstring}

\usepackage{cancel}
\newcommand{\xdownarrow}[1]{%
  {\left\downarrow\vbox to #1{}\right.\kern-\nulldelimiterspace}
}
  
\usepackage{cancel}

\begin{document}

\begin{titlepage}

\phantom{wowiezowie}

\vspace{-1cm}

\begin{center}

{\Huge {\bf 6d SCFTs and terminal singularities\\
\vspace{0.2cm}}}


\vspace{1cm}
{\Large Andrea Sangiovanni$^{\dagger\star\ddagger}$}\\ 

\vspace{1cm}

{\it
{\fontsize{9pt}{10pt}\selectfont
\renewcommand{\arraystretch}{2.5}

\begin{tabular}{c}
\makecell{$^\dagger$ \text{Mathematics Institute, Uppsala University,} \\ \text{Box 480, SE-75106 Uppsala, Sweden}}\\
\makecell{$^\star$ \text{Centre for Geometry and Physics, Uppsala University,} \\ \text{Box 480, SE-75106 Uppsala, Sweden}\\} \\
\makecell{$^\ddagger$ \text{Department of Physics and Astronomy, Uppsala University,}\\ \text{Box 516, SE-75120 Uppsala, Sweden}\\}\\
\end{tabular}
}}

\vskip .5cm
{\footnotesize \tt  andrea.sangiovanni@math.uu.se} \\

\vskip 1cm
     	{\bf Abstract }
\vskip .1in

\end{center}

\noindent 

The classification of 6d $\mathcal{N}=(1,0)$ SCFTs has been carried out with great success, enumerating consistent elliptically fibered threefolds in F-theory. In this note, we observe that the E-string 6d SCFT can be decorated with extra matter sectors. This decoration exploits an opening allowed by the current classification, implementing terminal singularities nested on the curve of the base of the elliptic fibration that supplies the tensor branch.  
Additionally, via dimensional reduction on a circle, we identify versions of the 5d $E_n$ theories decorated by extra matter sectors. As a bonus, the end of the resulting RG flow cascade reproduces a 5d theory recently constructed in the literature, explaining its 6d origin. We remain agnostic as to whether the decorated 6d and 5d theories constitute novel SCFTs.

\noindent

\eject

\end{titlepage}

\tableofcontents

\section{Introduction}\label{sec: intro}
Superconformal fixed points in spacetime dimension 6 were first discovered in \cite{Witten:1995ex,Strominger:1995ac,Ganor:1996mu,Seiberg:1996qx} via a string-theoretic approach. There exist only two superconformal algebras in six dimensions, respectively with 16 and 8 supercharges. 6d $\mathcal{N}=(2,0)$ superconformal field theories (SCFTs) admit a complete classification arising from F-theory on $\mathbb{C}^2/\Lambda \times \mathbb{T}^2$ (equivalently, from Type IIB string theory on $\mathbb{C}^2/\Lambda$), with $\Lambda \in ADE$, i.e.\ a finite subgroup of $SU(2)$ \cite{Witten:1995ex,Strominger:1995ac,Witten:1995zh}. The status of 6d $\mathcal{N}=(1,0)$ SCFTs is more subtle, and it requires a careful enumeration of consistent non-compact elliptically fibered Calabi-Yau threefolds (CY3) in an F-theory context, which take the form:
\begin{equation}
   \pi: \quad X \longrightarrow B, 
\end{equation}
where $B$ is a non-compact complex surface. At the conformal point, $B$ is an orbifold $\mathbb{C}^2/\Gamma$, with $\Gamma$ specific finite subgroups of $U(2)$ \cite{Heckman:2013pva}. On the tensor branch in the corresponding Type IIB description, D3-branes wrapping compact 2-cycles in the base give rise to strings, that become tensionless once the volume of such cycles is sent to zero, reaching the superconformal fixed point.
The shrinkability of the base curves imposes strict geometric constraints: each curve should be isomorphic to $\mathbb{P}^1$, and their intersection pairing should be negative definite. This guarantees that each curve has negative self-intersection, and that all the curves can be shrunk simultaneously to reach the fixed point. Then, suitable degenerations of the fibers must be chosen so that the whole threefold is an elliptically fibered CY3. The severity of the singularities of the fibers on top of curves in the base dictates the corresponding gauge algebra, and charged matter arises at codimension 2 points in the base, where discriminant components collide and the elliptic fibers undergo further degeneration \cite{KatzVafa}. \\
\indent Relying on this approach, the tensor branches of 6d theories which admit a SCFT completion have been classified in \cite{Heckman:2013pva,Heckman:2015bfa}, showing that only a finite set of fundamental building blocks is needed in order to capture all geometric 6d SCFTs. These are the non-Higgsable clusters \cite{Morrison:2012np}, ADE-shaped configurations of curves with self-intersection $-2$, and a curve with self-intersection $-1$ and global symmetry $E_8$, known as the E-string\footnote{Alternatively, the E-string can be constructed as the theory on the worldvolume of an M5-brane on top of an $E_8$ wall, with M2-branes stretching between the M5 and the wall \cite{Ganor:1996mu}.}.\\

\indent In this work we explore a class of threefold geometries that engineer 6d $\mathcal{N}=(1,0)$ theories via F-theory, and that are not captured by the current classification scheme. We observe that the E-string 6d SCFT can be decorated with a codimension 3 singularity in the threefold, sitting on top of the $-1$ curve in the tensor branch phase. This decoration does not change the self-intersection of the curve, it does not change the (absent) gauge algebra supported by the curve, and it does not change the (absent) gauge-charged matter: hence, it is invisible to the classification data of \cite{Heckman:2013pva,Heckman:2015bfa}. The decoration does, however, change the spectrum on the tensor branch: over a point of the $-1$ curve, the threefold has a terminal singularity
(specifically, of the family known as ``Reid's pagodas''), which supplies extra matter localized on the $-1$ curve of the E-string\footnote{We stress that the decoration is engineered by terms of higher degree in the Weierstrass data: in the blown-down geometry, these irrelevant terms are argued in \cite{Heckman:2013pva} not to affect the fixed point. This is compatible with the agnostic stance we take below: the decoration changes the geometry and the tensor-branch spectrum, while whether it changes the fixed point is an open question.}. Terminal singularities admit at most a small resolution (namely, they admit no crepant resolution with exceptional compact divisors), and have been studied in the context of F-theory in compact settings in \cite{Arras:2016evy,Grassi:2018rva}.\\
\indent We explicitly construct a Weierstrass model encoding the decorated E-string, which comes in a family of theories labelled by the terminal singularities hosted on the $-1$ curve. Then, via dimensional reduction on a circle, we show that the decorated E-string descends to decorated versions of the $E_n$ 5d SCFTs, $n=1,\ldots,8$: the bottom entry of the RG flow obtained by successively decoupling matter from the 5d KK theory is a decorated version of the $E_1$ 5d SCFT, recently studied in detail in \cite{Collinucci:2025rrh} (see also \cite{Dramburg:2025tlb} for an analysis of its defect group). Circle-reduction serves a two-fold purpose. First, it shows that the extra matter provided by the terminal singularity in 6d does not increase the \textit{geometric} flavor symmetry rank. This is compatible with two interpretations:
\begin{itemize}
    \item the matter sector supplied by the terminal singularity is coupled to the E-string SCFT, and carries no extra flavor symmetry beyond that of the ordinary E-string;
    \item the matter sector supplied by the terminal singularity is \textit{not} coupled to the E-string SCFT: both the extra matter and the SCFT are simply charged under a shared background flavor symmetry, thus not increasing its rank. 
\end{itemize}
We remain agnostic in this respect, and clearly an independent analysis of the status of this theory is desirable in future work.  Second, the analysis of the circle reduction of the decorated E-string furnishes an explicit 6d parent for the theory of \cite{Collinucci:2025rrh}, confirming once again the healthy status of the conjecture arguing that all 5d SCFTs should enjoy an origin from 6d \cite{Jefferson_2018,Bhardwaj:2019xeg}. Altogether, our task here is to show that the gap observed in 5d \cite{Closset:2020scj,Collinucci:2025rrh}, namely the classification schemes missing decorations of SCFTs by terminal singularities, \textit{is present in a similar form in 6d}.

\subsection*{Outline}
In Section \ref{sec: E string} we review a handful of basic F-theory concepts, including the realization of the E-string via an elliptically fibered CY3. We then refine the construction in Section \ref{sec: dec E string} introducing a terminal singularity on top of the curve of the E-string in the base, hosting an extra matter sector. The dimensional reduction of the decorated E-string theory is studied in Section \ref{sec: 5d SCFTs}, showing that its descendants are decorated $E_n$ 5d SCFTs. In Section \ref{sec: outlook} we collect some concluding remarks.

\section{E-string from F-theory: a short review}\label{sec: E string}
In order to construct 6d $\mathcal{N}=(1,0)$ SCFTs we employ the standard setup relying on F-theory geometric engineering on an elliptically fibered CY3 $X$ \cite{Vafa:1996xn}. Concretely, $X$ can be presented via a Weierstrass model:
\begin{equation}\label{Weierstrass model}
  X: \quad  y^2 = x^3+fxz^4+gz^6,
\end{equation}
where $x,y,z$ are coordinates on the weighted projective space $\mathbb{P}_{231}$ (we label the related $\mathbb{C}^*$ action as $\mathbb{C}^*_1$). $x,y,z$ and $f,g$ are appropriate sections of powers of a line bundle $L$ over the two-complex-dimensional base $B$, that encodes the variation of the axiodilaton over the base:
\begin{equation}\label{line bundle}
    x \in \Gamma(B,L^2), \quad y \in \Gamma(B,L^3), \quad z \in \Gamma(B,\mathcal{O}), \quad f \in \Gamma(B,L^4), \quad g \in \Gamma(B,L^6),
\end{equation}
with $\mathcal{O}$ the trivial line bundle. The Calabi-Yau condition for the threefold is enforced by imposing $L = -K_B$, guaranteeing that the (non-trivial) first Chern class of the base is cancelled by that of the fiber. The discriminant, which describes the location of 7-branes in the Type IIB description, can be read off as:
\begin{equation}
    \Delta = 4f^3+27g^2.
\end{equation}
Under appropriate conditions \cite{Katz:2011qp} (always satisfied for the cases discussed in this work), the Weierstrass model can equivalently be described by a threefold in Tate form:
\begin{equation}\label{tate model}
    X_T: \quad y^2+a_1xyz+a_3yz^3=x^3+a_2x^2z^2+a_4xz^4+a_6z^6,
\end{equation}
where the parameters $a_i$ are globally defined sections of degree $i$ of the line bundle $L$. The Weierstrass model description can be retrieved by defining:
\begin{equation}
    b_2 = a_1^2+4a_2, \quad b_4 = a_1a_3+2a_4, \quad b_6 = a_3^2+4a_6,
\end{equation}
and then obtaining the sections $f$ and $g$:
\begin{equation}
    f = -\frac{1}{48}(b_2^2-24b_4), \quad g = -\frac{1}{864}(-b_2^3+36b_2b_4-216b_6).
\end{equation}
We now review the construction of the E-string SCFT, employing the Tate model \eqref{tate model}. On the tensor branch, the base contains a single non-trivial class in homology, represented by a curve $C$ isomorphic to $\mathbb{P}^1$ and with self-intersection equal to $-1$. Concretely, we realize the resolved base as:
\begin{equation}\label{base}
    B =\left( \mathbb{C}^3_{s_1,s_2,\tau} \smallsetminus \{s_1=s_2=0\} \right)/\mathbb{C}^*_2,
\end{equation}
where $s_1,s_2$ have charge 1 under the $\mathbb{C}^*$-action (that we label as $\mathbb{C}^*_2$), and $\tau$ has charge $-1$. Hence $B$ is simply the total space of the line bundle $\mathcal{O}(-1)$ over a $\mathbb{P}^1$:
\begin{equation}\label{P1 fibration}
   B= \text{Tot}\,\mathcal{O}_{\mathbb{P}^1}(-1).
\end{equation}
The CY3 over the resolved base $B$ is then defined by the hypersurface \eqref{tate model} in a toric ambient space, where the $a_i$ are polynomials of appropriate degree in $s_1,s_2,\tau$:
\begin{equation}\label{toric ambient space}
    \begin{array}{ccccccc}
         & x & y & z & s_1 & s_2 & \tau \\
         \hline
        \mathbb{C}^*_1 & 2 & 3 & 1 & 0 & 0 & 0 \\
        \mathbb{C}^*_2 & 2 & 3 & 0 & 1 & 1 & -1 \\
    \end{array}
\end{equation}
The origin of the tensor branch, namely the E-string SCFT, is reached via the blowdown map:
\begin{equation}
    (x_1,x_2) = (s_1\tau,s_2\tau).
\end{equation}
The curve in the base $B$ supporting the E-string is the curve $C$ defined by the locus $\tau = 0$, which is the exceptional locus of the blowup:
\begin{equation}
    C = \{ \tau = 0\} \cong \mathbb{P}^1_{[s_1:s_2]} \subset B.
\end{equation}
We can check its self-intersection in the base explicitly, using \eqref{P1 fibration}:
\begin{equation}
    C^2 = \text{deg}(\mathcal{O}_{\mathbb{P}^1}(-1)) = -1,
\end{equation}
since $C$ is the zero-section of the $\mathcal{O}_{\mathbb{P}^1}(-1)$ line bundle. By adjunction we also get (denoting as $K_B$ the canonical line bundle of the base):
\begin{equation}
    -K_{B}\cdot C = C^2+2 = 1.
\end{equation}
Restricting the discriminant to the curve $C$ we get:
\begin{equation}\label{Delta}
    \text{deg}(\Delta|_C) = -12K_B \cdot C = 12.
\end{equation}
For generic $a_i$ \eqref{Delta} implies that $\Delta|_C$ vanishes at order 1 on twelve distinct points. Since $f$ and $g$ are non-zero on those points we get twelve special fibers with vanishing orders:
\begin{equation}
    \text{ord}(f,g,\Delta) = (0,0,1),
\end{equation}
i.e.\ twelve fibers of type $I_1$. This is the hallmark of the usual E-string SCFT, which is recovered by contracting the curve $C=\{\tau = 0\}$ in the base $B$ \eqref{base}.

\section{The decorated E-string and terminal singularities}\label{sec: dec E string}
In this section we construct a decorated version of the E-string, in which a codimension 2 point in the base supports a terminal threefold singularity, locally of the form:
\begin{equation}\label{reid pagoda}
    w_1^2+w_2^2+w_3^2+w_4^{2p} = 0,
\end{equation}
which is also known as Reid's pagoda threefold \cite{reid1983minimal}, for $p\geq 1$. Notice that $p=1$ is simply the conifold threefold. The presence of a singular locus of type \eqref{reid pagoda} provides an extra matter sector on top of the E-string content. This can be neatly seen by noticing that \eqref{reid pagoda} only admits a small resolution which inflates a single $\mathbb{P}^1$, with associated Gopakumar-Vafa invariant $n_{[\mathbb{P}^1]}^{g=0}=p$. In the dual M-theory geometric engineering picture, this amounts to a trivial Coulomb branch (given the absence of compact divisors in the crepant resolution), and a Higgs branch with $p$ quaternionic dimensions. We refer to \cite{Collinucci:2021wty,Collinucci:2021ofd} for further details on its features. The degrees of freedom that span the Higgs branch are the extra matter sector that decorates the E-string in the 6d perspective adopted here\footnote{It is not clear whether Reid's pagoda is a theory of free hypermultiplets, or a more involved theory. The ascertained geometric facts are the presence of a $U(1)$ flavor symmetry due to the small resolution and of a Higgs branch of quaternionic dimension $p$ ($p$ free hypermultiplets would enjoy a $Sp(p)$ flavor symmetry, which however need not be geometrically manifest). The mismatch between geometric flavor symmetry and Higgs branch dimension is ubiquitous for terminal threefolds. For this reason, we will generically denote the decoration produced by Reid's pagoda terminal threefold as ``extra matter''.}.\\

We wish to engineer this matter locus along $a_3=a_4=0$ in the Tate model \eqref{tate model}. In order to realize it explicitly, we consider the following global choice of parameters:
\begin{equation}\label{parameters choice}
    a_1 = 0, \quad a_2 = c_1s_2^2,\quad a_3= s_1 s_2^2, \quad a_4 = s_1s_2^3+c_2 s_1^4+c_3\tau^p s_2^{p+4}, \quad a_6 = 0,
\end{equation}
with $c_1,c_2,c_3$ generic constant coefficients ($c_1\neq 0,1$ and $c_2,c_3\neq 0$), and $s_1,s_2,\tau$ the coordinates on the blown-up base $\mathcal{O}(-1)\rightarrow \mathbb{P}^1$, exactly as in the usual E-string case reviewed in the previous section. We keep denoting by $C=\{\tau=0\} \subset B$ the exceptional curve of the blowup of the base.
This gives rise to the Tate model:
\begin{equation}\label{pagoda tate}
  X_P: \quad  y^2+(s_1s_2^2)yz^3=x^3+(c_1s_2^2)x^2z^2+(s_1s_2^3+c_2s_1^4+c_3\tau^p s_2^{p+4})xz^4,
\end{equation}
whose singular locus is a point on top of $C$ (for generic values of the $c_i$):
\begin{equation}\label{singular point}
    x=y=s_1=\tau = 0.
\end{equation}
We will later prove that the singular point \eqref{singular point} is precisely of the pagoda form \eqref{reid pagoda}.

The choice \eqref{parameters choice} is highly constrained:
\begin{itemize}
    \item $a_6 = 0$ is famously known as the $U(1)$ restriction \cite{Grimm:2010ez,Braun:2011zm,Lee:2018ihr}, which guarantees the appearance of an extra section in the elliptic threefold (other than the usual $S_0 = \{z=0=y^2-x^3\}$), corresponding to a non-trivial rank 1 Mordell-Weil group. This can be already seen in the general Tate model \eqref{tate model}, which upon setting $a_6=0$ becomes:
    \begin{equation}
        y(y+a_1xz+a_3z^3)=x(x^2+a_2xz^2+a_4z^4).
    \end{equation}
    In this form, the extra section $S=\{x=0=y\}$ becomes visible: as we will see in Section \ref{sec: resolution}, in our model $S$ passes through the singular point \eqref{singular point}, and blowing up along $S$ resolves it.
    \item $a_3 = s_1s_2^2$: since we wish to engineer a matter locus on $a_3=a_4=0$, then $a_3$ must vanish on the point \eqref{singular point}. This is achieved by introducing a dependence on $s_1$, and padding it with enough powers of $s_2$ so as to reach the correct charge.
    \item $a_4 = s_1s_2^3+c_2 s_1^4+c_3\tau^p s_2^{p+4}$: this coefficient is fixed by requiring that it vanishes on the point \eqref{singular point} (hence all terms depend either on $s_1$ or $\tau$), that the germ of Reid's pagoda is reproduced (guaranteed by $c_3\tau^p s_2^{p+4}$, which has the correct charge for every $p$), and that there is no non-compact curve supporting a flavor algebra: hence the presence of the $c_2s_1^4$ term, that ensures that there is no common $s_2$ factor, which would produce a non-compact curve in the base supporting a singular fiber. Conversely, for $c_3=0$ both $a_3$ and $a_4$ would be divisible by $s_1$, and the threefold would be singular along the whole non-compact curve $x=y=s_1=0$, i.e.\ an $I_2$ locus supporting an $\mathfrak{su}(2)$ flavor algebra (we return to this point at the end of this section).
    \item $a_2=c_1s_2^2$: the only requirement is that $a_2$ does not vanish at the singular point \eqref{singular point}, where $a_2|_{s_1=\tau=0}=c_1s_2^2$ with $c_1\neq0$ (and also $c_1\neq 1$, as we will see below \eqref{delta on C}). This guarantees that the fiber over the singular point is of type $I_2$. Further terms of charge 2 (such as $s_1s_2$, $s_1^2$, $\tau s_2^3$) can be added without changing the singularity, since under the scaling \eqref{scaling} they only contribute terms of higher weight that do not perturb the germ.
    \item $a_1=0$ is merely a simplification choice.
\end{itemize}
With the Tate model \eqref{pagoda tate} in hand, we can compute $f,g,\Delta$ of the corresponding Weierstrass model. The result is:
\begin{equation}
\scalemath{0.9}{
    \begin{split}
  &      f = \frac{1}{48} \left(-16 c_1^2 s_2^4+48 \left(c_2 s_1^4+s_1
   s_2^3+c_3 s_2^{4+p} \tau ^p\right)\right),\\
 &  g = \frac{1}{864} \left(216 s_1^2 s_2^4+64 c_1^3 s_2^6-288 c_1
   s_2^2 \left(c_2 s_1^4+s_1 s_2^3+c_3 s_2^{4+p} \tau
   ^p\right)\right),\\
   & \Delta = \frac{\left(-16 c_1^2 s_2^4+48 \left(c_2 s_1^4+s_1
   s_2^3+c_3 s_2^{4+p} \tau ^p\right)\right)^3+64 s_2^4 \left(27
   s_1^2 s_2^2+8 c_1^3 s_2^4-36 c_1 \left(c_2
   s_1^4+s_1 s_2^3+c_3 s_2^{4+p} \tau ^p\right)\right)^2}{27648}.\\
    \end{split}}
\end{equation}
Evaluating $\Delta$ on the curve $C = \{\tau = 0\}$, we get:
\begin{equation}\label{Delta tuned}
\scalemath{0.9}{
\begin{split}
    \Delta|_C &= \frac{s_1^2}{16} \left(64 c_2^3 s_1^{10}+192 c_2^2 s_1^7 s_2^3-16
   c_1^2 c_2^2 s_1^6 s_2^4+192 c_2 s_1^4 s_2^6-72
   c_1 c_2 s_1^4 s_2^6-32 c_1^2 c_2 s_1^3
   s_2^7+27 s_1^2 s_2^8\right.\\
   &\left.+64 s_1 s_2^9-72 c_1 s_1
   s_2^9-16 c_1^2 s_2^{10}+16 c_1^3 s_2^{10}\right).
   \end{split}}
\end{equation}
From \eqref{Delta tuned} we see that the fiber over $C$ at the point $s_1=0$ has vanishing orders
\begin{equation}\label{delta on C}
    \text{ord}(f,g,\Delta)|_C = (0,0,2),
\end{equation}
namely a fiber of type $I_2$. Indeed, at $s_1=0$ the bracket in \eqref{Delta tuned} reduces to $16c_1^2(c_1-1)s_2^{10}$, which is why we require $c_1\neq0,1$: for $c_1=0$ also $f$ and $g$ vanish at this point and the fiber is no longer of type $I_2$, while for $c_1=1$ the vanishing order of $\Delta$ jumps.\\
\indent This is the hallmark of the decorated E-string model: two of the twelve $I_1$ fibers of the ordinary E-string have collided, producing an $I_2$ fiber, which sits precisely on top of the intersection $a_3=a_4=0$, where a terminal singularity resides, for every value of $p$. We can now directly inspect the germ of the Tate model \eqref{pagoda tate} at its singular point \eqref{singular point}: to this end, we can safely set $z = s_2 = 1$, working in a patch near the singularity. In order to detect the form \eqref{reid pagoda} explicitly, we employ appropriate coordinates, using the change of variables (this holds for $c_1 \neq 1$):
\begin{equation}
    x =\zeta- \frac{c_3}{2(c_1-1)}\tau^p, \quad y=u+\zeta- \frac{c_3}{2(c_1-1)}\tau^p, \quad s_1 = v-u-2\zeta+\frac{c_3}{(c_1-1)}\tau^p.
\end{equation}
The resulting threefold is \textit{not} quasi-homogeneous. To identify the germ, we introduce a scaling action on the coordinates:
\begin{equation}\label{scaling}
    u\rightarrow \lambda^{1/2}u,\quad  v\rightarrow \lambda^{1/2}v,\quad  \zeta\rightarrow \lambda^{1/2}\zeta,\quad  \tau\rightarrow \lambda^{1/(2p)}\tau.
\end{equation}
In this fashion, we see that the only terms that scale as $\lambda$ are:
\begin{equation}\label{pagoda patch}
    uv = (c_1-1)\zeta^2-\frac{c_3^2}{4(c_1-1)}\tau^{2p},
\end{equation}
while all the rest scale as higher powers of $\lambda$, and are hence irrelevant in the vicinity of the singularity. Clearly, upon suitable renaming and rescaling of the variables, the singularity germ \eqref{pagoda patch} is identical to Reid's pagoda singularity displayed in \eqref{reid pagoda}. This shows that the F-theory model \eqref{pagoda tate} is engineering an E-string-like theory with a Reid pagoda singularity nested on the $-1$ curve which arises from the tensor branch. As can be seen from the dual M-theory perspective \cite{Collinucci:2021wty,Collinucci:2021ofd}\footnote{Or, in principle, by embedding Reid's pagoda singularity into a compact elliptically fibered threefold employing the same techniques of \cite{Arras:2016evy}, and computing the complex structure deformations of that threefold. The dimension of the Higgs branch engineered by Reid's pagoda is related to the Milnor number of the singularity $\mu$. Namely: $p= \text{dim}_{\mathbb{H}}\,(\rm HB) =\frac{\mu+1}{2}$, because of the presence of a single $U(1)$ arising from the crepant resolution. We refer to \cite{Closset:2020scj} for further details on the general relation between dimension of the Higgs branch and Milnor number of isolated singularities.}, the Reid pagoda singularity supports a $p$-dimensional Higgs branch arising from M2-branes wrapped on the exceptional $\mathbb{P}^1$ inflated by its resolution, which become massless in the singular limit. We will show the explicit resolution in Section \ref{sec: resolution}, observing that the extra matter is charged under the flavor $U(1)$ associated with the extra section.\\
\indent The role of $c_3$ is now transparent. For $c_3=0$, the germ \eqref{pagoda patch} reduces to
\begin{equation}
    uv=(c_1-1)\zeta^2,
\end{equation}
which is an $A_1$ singularity fibered over the $\tau$-line, i.e.\ the local form of the $I_2$ flavor locus $\{s_1=0\}$ mentioned in the bullet on $a_4$. Turning on $c_3$ gives:
\begin{equation}
    uv=\det\big(\sqrt{c_1-1}\,\zeta\,\mathbbm{1}_2-\Phi\big),
\end{equation}
with:
\begin{equation}\label{higgs bckg}
   \Phi=\kappa\begin{pmatrix}
       \tau^p & 0 \\
       0 & -\tau^p \\
   \end{pmatrix}, \quad \kappa^2=c_3^2/(4(c_1-1)).
\end{equation}
\eqref{higgs bckg} is a position-dependent Higgs background on the 8d theory supported on the flavor brane \cite{Cecotti:2010bp}, which localizes the matter at $\tau=0$. This is the 6d counterpart of the non-constant flavor backgrounds of \cite{Collinucci:2025rrh}.

\subsection{Small resolution of the terminal singularity}\label{sec: resolution}
In this section we explicitly construct the small resolution that takes care of smoothing out the nested terminal singularity on the E-string. 

The resolution can be performed in the toric ambient space \eqref{toric ambient space}, namely blowing up $X_P$ along the Weil non-Cartier divisor $S=\{x=y=0\}$:
\begin{equation}\label{blowup map}
    x \rightarrow \tilde{x}\delta, \quad y\rightarrow \tilde{y}\delta.
\end{equation}
This produces a small resolution  of the Tate model $X_P$ in \eqref{pagoda tate}:
\begin{equation}
    \tilde{X}_P \longrightarrow X_P.
\end{equation}
Concretely, the resolution produces the proper transform:
\begin{equation}\label{resolved tate}
    \tilde{X}_P: \quad \delta\tilde{y}^2+(s_1s_2^2)\tilde{y}z^3=\delta^2\tilde{x}^3+(c_1s_2^2)\delta\tilde{x}^2z^2+(s_1s_2^3+c_2s_1^4+c_3\tau^p s_2^{p+4})\tilde{x}z^4,
\end{equation}
which is completely smooth. \eqref{resolved tate} is a hypersurface in the toric ambient space:
\begin{equation}\label{toric ambient space 2}
    \begin{array}{cccccccc}
         & \tilde{x} & \tilde{y} & z & s_1 & s_2 & \tau & \delta \\
         \hline
        \mathbb{C}^*_1 & 2 & 3 & 1 & 0 & 0 & 0 & 0 \\
        \mathbb{C}^*_2 & 2 & 3 & 0 & 1 & 1 & -1 & 0 \\
        \mathbb{C}^*_3 & 1 & 1 & 0 & 0 & 0 & 0 & -1\\
    \end{array}
\end{equation}
The SR ideal reads:
\begin{equation}
    \{s_1=s_2=0\},\quad \{\tilde{x}=\tilde{y}=0\},\quad \{\delta=z=0\}.
\end{equation}
As a consequence of the blowup, the extra section $x=y=0$ arising from the $U(1)$ restriction now corresponds to a Cartier divisor. 
Setting $\delta=0$ in \eqref{resolved tate} gives the proper transform of the section $\tilde{S}$. Its fiber over the singular point \eqref{singular point} is the exceptional curve of the small resolution, which we denote as $b$:
\begin{equation}\label{b equation}
    b:\quad s_1 = \tau = \delta = 0 \quad \cong \quad \mathbb{P}^1_{[\tilde{x}:\tilde{y}]}.
\end{equation}
On top of the singular point lies also another curve $b'$, as can be seen by imposing $s_1=\tau = 0$ into \eqref{resolved tate}:
\begin{equation}
    b': \quad s_1 = \tau = \tilde{y}^2-\delta\tilde{x}^3-c_1s_2^2\tilde{x}^2z^2=0.
\end{equation}
The curve $b'$ is simply the proper transform of the nodal fiber on top of the singular point, and it intersects $b$ in two distinct points, compatibly with the fact that this is a $I_2$ fiber:
\begin{equation}\label{intersections}
b \cap b': \quad   \delta= s_1 =\tau = \tilde{y}\pm \sqrt{c_1}s_2\tilde{x}z=0.
\end{equation}

\indent We can check the charge of the states wrapping $b$ under the global $U(1)$ symmetry generated by the small resolution. To this end, we need to identify the image of the resolved section $\tilde{S}$ (the proper transform of $S$), that has been turned into a Cartier divisor after the small resolution, under the Shioda map $\sigma$ \cite{Shioda,Morrison:2012ei}. $\sigma(\tilde{S})$ should satisfy appropriate transversality conditions, enforcing vanishing intersection with curves in the base, so that it does not mix with the 6d tensors, and vanishing intersection with the generic elliptic fiber, so as to decouple it from the KK $U(1)$. These are sufficient conditions in our setup (given the absence of W-bosons charged under some gauge algebra) to produce the ansatz (following the conventions of \cite{weig2018tasi}):
\begin{equation}
    \sigma(\tilde{S}) = \tilde{S}-S_0 -\pi^{-1}(\pi_*((\tilde{S}-S_0)\cdot S_0)),
\end{equation}
with $\pi_*$ the push-forward along the elliptic fibration. Recall that in our geometry $\pi_*(\tilde{S}\cdot S_0) = 0$ because of the SR ideal, and $\pi_*(S_0\cdot S_0) = K_B$ by adjunction, as:
\begin{equation}
    \mathcal{O}(S_0)|_{S_0} =N_{S_0/\tilde{X}_P} = K_{S_0} \cong K_B,
\end{equation}
which holds since $S_0$ is isomorphic to $B$. Then we see that:
\begin{equation}
     \sigma(\tilde{S}) = \tilde{S}-S_0 +\pi^{-1}(K_B).
\end{equation}
$\tilde{S}$ is given by $\delta=0$ intersected with the blown-up threefold equation. Notice that using the SR ideal one can easily check that (recalling $S_0 = \{z=0\}$):
\begin{equation}\label{b intersections}
    b \cdot S_0 = 0, \quad b' \cdot S_0 = 1.
\end{equation}
Analogously:
\begin{equation}\label{b Stilde}
    b \cdot \tilde{S} = -1, \quad b'\cdot \tilde{S} = 2.
\end{equation}
In particular, $b'\cdot \tilde{S} = 2$ because of \eqref{intersections}. $b \cdot\tilde{S} =-1$ can be proven noticing that $s_2,z\neq 0$ on $b$ \eqref{b equation} because of the SR ideal. Then we can use $\mathbb{C}_1^*$ to set $z=1$ and $\mathbb{C}_2^*$ to set $s_2=1$. We are then left only with the action of $\mathbb{C}_3^*$ on $(\tilde{x},\tilde{y},\delta)$:
\begin{equation}
    (\tilde{x},\tilde{y},\delta) \sim (t\tilde{x},t\tilde{y},t^{-1}\delta).
\end{equation}
Near $b$ the ambient space is thus $\text{Tot}\,\mathcal{O}_{\mathbb{P}^1}(-1)\times \mathbb{C}^2_{s_1,\tau}$ with $b$ the $\mathbb{P}^1_{[\tilde{x}:\tilde{y}]}$ at $s_1=\tau=0$ and $\tilde{S}=\{\delta=0\}$ the zero section. Since $\delta$ has $\mathbb{C}_3^*$ charge $-1$, then $\mathcal{O}(\tilde{S})|_b = \mathcal{O}_{\mathbb{P}^1}(-1)$, giving $\tilde{S}\cdot b =-1$ in \eqref{b Stilde}, for every value of $p$.
 Collecting all the intersection data, the charge $q_b$ of $b$ under the flavor $U(1)$ is given by (using the projection formula $\pi^{-1}K_B\cdot b = K_B\cdot \pi_* b$, and $\pi_* b = 0$):
\begin{equation}\label{charges}
    q_b = \sigma(\tilde{S})\cdot b = \tilde{S}\cdot b -S_0\cdot b+\pi^{-1}(K_B)\cdot b = -1+0+0 = -1.
\end{equation}
\eqref{charges} shows that the extra matter engineered by M2-branes wrapping the curve $b$ (in the M-theory picture) has charge $\pm 1$ (depending on the wrapping orientation) under the flavor $U(1)$ related to the extra section of our model \eqref{pagoda tate}. As a consistency check, we compute:
\begin{equation}
    q_{b'}=\sigma(\tilde S)\cdot b'=2-1+0=1,
\end{equation}
so that $q_b+q_{b'}=\sigma(\tilde S)\cdot F=0$, as required for the full fiber.

\section{5d SCFTs from circle reduction of the decorated E-string}\label{sec: 5d SCFTs}
In this section, we explore the circle reduction of the decorated E-string theory constructed above. We will show two facts, fleshed out in Sections \ref{sec: flavor rank} and \ref{sec: RG cascade} respectively:
\begin{itemize}
\item The 5d KK theory after circle reduction is engineered by M-theory on $\tilde X_P$, whose only compact divisor is a rational elliptic surface $dP_9$\footnote{Following the widespread notation of the physics literature, we denote by $dP_9$ the rational elliptic surface, i.e.\ $\mathbb P^2$ blown up at the nine base points. Strictly speaking, it is not a del Pezzo surface: $-K$ is nef, but $K^2=0$.} containing the exceptional curve $b$ arising from the resolution of the terminal singularity. Its geometric flavor rank is $r_{\text{flavor}}=9$ for every $p$, exactly as for the circle reduction of the ordinary E-string: this is compatible both with the hypothesis that the extra matter supplied by the terminal singularity is coupled to the E-string, and with the possibility that it is simply decoupled extra matter;
  \item Starting from the KK theory engineered by $\tilde{X}_P$, we can decouple matter fields by flopping curves out of the compact divisor $dP_9$: this produces a cascade of genuine 5d theories which are decorated versions of the 5d $E_n$ theories with $n=1,\ldots,8$, with extra matter supplied by the terminal singularity. The case $n=1$, which is found at the end of the descent, is the theory engineered via orbifold techniques in \cite{Collinucci:2025rrh}.
\end{itemize}

Consider F-theory on $X_P$ \eqref{pagoda tate} on the background $S^1_R \times \mathbb{R}^{1,4}$, dual to M-theory on the resolved threefold $\tilde{X}_P$ in \eqref{resolved tate}. The radius $R$ is inversely proportional to the volume of the elliptic fiber: in the F-theory limit the volume of the fiber is sent to zero, recovering the 6d theory, while at finite radius a 5d KK theory is obtained. The threefold geometry $\tilde{X}_P$ only has one compact divisor $D$, which is the pull-back of the base curve:
\begin{equation}\label{divisor smooth}
    D = \pi^{-1}(C).
\end{equation}
By adjunction (and recalling that $K_{\tilde{X}_P}=0$, since the resolved threefold is Calabi-Yau) we get:
\begin{equation}\label{KD adjunction}
    K_D = (K_{\tilde{X}_P}+ D)|_D = D|_D = \pi^{-1}(N_{C/B}) = \pi^{-1}(\mathcal{O}_{\mathbb{P}^1}(C^2)) = \pi^{-1}(\mathcal{O}_{\mathbb{P}^1}(-1)).
\end{equation}
Notice that $\pi^{-1}(\mathcal{O}_{\mathbb{P}^1}(-1)) = -F$, with $F$ the fiber class, since the pullback of the point class of $C$ along the threefold is an elliptic fiber. Hence we obtain:
\begin{equation}\label{K fiber}
    K_D = -F.
\end{equation}
In order to identify the compact divisor $D$ we compute the square of its canonical bundle, which is:
\begin{equation}
    K_D^2 = (-F)^2 = 0,
\end{equation}
since two distinct fibers do not intersect. Moreover, we have that $-K_B\cdot C=1$, and the restrictions of $f$ and $g$ to $C\simeq\mathbb P^1$ have degrees $4$ and $6$. Hence $D$ is a smooth elliptic surface over $\mathbb P^1$ whose Weierstrass model has $\deg(f,g)=(4,6)$, i.e.\ the rational elliptic surface dP$_9$, exactly as in the circle reduction for the ordinary E-string\footnote{When the curve $b$ is shrunk to zero size, retrieving the singular Tate model \eqref{pagoda tate}, the divisor acquires a $A_1$ singular point where the terminal singularity of the full threefold is nested.}.\\
\indent The crucial difference between the ordinary and the decorated E-string is that in the latter case the curve $b$, resulting from the blowup of the terminal singularity, is contained in $D$. Note that the small resolution only adds $b$: all the other curves of $D$ (e.g.\ the sections of $D\to C$) are already present in $X_P$, as for the ordinary E-string.\\
\indent To explore the properties of the divisor $D$, we choose a basis of its Picard group $\text{Pic}(D)$:
\begin{equation}\label{basis}
\begin{split}
    & H, E_i, i=1,\ldots, 9, \quad \text{with: } H^2 = 1, \quad E_i\cdot E_j = -\delta_{ij},\\
    & H\cdot E_i = 0, \quad K_D = -3H+\sum_iE_i.
    \end{split}
\end{equation}
Since the curve $b$ supporting the extra matter lies in a fiber, we have:
\begin{equation}\label{b fiber}
    b\cdot F = 0.
\end{equation}
Moreover, using adjunction:
\begin{equation}\label{b square}
    b^2 = (2g-2)-K_D\cdot b = -2+0 = -2,
\end{equation}
where we have used \eqref{K fiber} and \eqref{b fiber}. Hence $b$ is a root in the lattice of curves orthogonal to $K_D$.

\subsection{The flavor rank and decoupled/coupled extra matter}\label{sec: flavor rank}

Everything is now in place to discuss the flavor rank of the 5d theory, and its implications for the relationship between the extra matter of the decorated E-string and the SCFT sector. The flavor rank $r_{\text{flavor}}$ of the 5d KK theory depends only on a neighborhood of the compact divisor, and is counted by classes in $H_2(\tilde X_P,\mathbb R)$ that are not paired to classes in $H_4(\tilde X_P,\mathbb R)$:
\begin{equation}\label{flavor rank}
r_{\text{flavor}}=\dim H_2(\tilde X_P,\mathbb R)-\dim H_4(\tilde X_P,\mathbb R)\,,
\end{equation}
with $H_4$ generated by $D$, and $H_2$ generated by the curves in $D$. Since $b\subset D$, it does not add a new class, and
\begin{equation}\label{flavor rank value}
r_{\text{flavor}}=9
\end{equation}
for every value of $p$, as for the circle reduction of the ordinary E-string. This alone does not tell whether the extra matter is coupled to the E-string. Indeed, if it were decoupled, the theory would simply be a tensor product:
\begin{equation}\label{decoupled theory}
\mathcal T_{\rm decoupled}=\text{E-string}\otimes\text{extra matter}\,.
\end{equation}
In this case one might expect that the flavor rank is:
\begin{equation}\label{decoupled flavor}
r_{\text{flavor}}^{\rm decoupled}=9+n_p\,,
\end{equation}
with $n_p>0$ the rank of the flavor symmetry of the extra matter. This, however, assumes that the geometry provides an independent mass parameter for the extra matter, which need not be the case. The extra matter arises from M2-branes on $b$, so its mass is ${\rm vol}(b)$. By \eqref{b intersections} and \eqref{b fiber}, $b$ is orthogonal both to the zero section $Z\equiv S_0\cap D$ of $D$ and to the fiber:
\begin{equation}
b\cdot Z=0\,,\qquad b\cdot F=0\,.
\end{equation}
The $E_8$ lattice of $dP_9$ is precisely the orthogonal complement of $Z$ and $F$ in ${\rm Pic}(D)$:
\begin{equation}
\Lambda_{E_8}=\{x\in{\rm Pic}(D)\ :\ x\cdot Z=x\cdot F=0\}\,.
\end{equation}
Hence $b\in\Lambda_{E_8}$, and ${\rm vol}(b)$ is a combination of the $E_8$ mass parameters only. The same holds for the $U(1)$ arising from the extra section of our $U(1)$-restricted model: by construction of the Shioda map, $\sigma(\tilde S)$ is orthogonal to $F$ and to curves in the zero section, so that
\begin{equation}
\sigma(\tilde S)|_D\in\Lambda_{E_8}\,,\qquad \sigma(\tilde S)\cdot b=-1\neq0\,.
\end{equation}
Therefore this $U(1)$ is a Cartan $U(1)$ of $E_8$, acting on the E-string sector as well as on the extra matter. In a tensor product of the form \eqref{decoupled theory}, it would be the diagonal combination of a Cartan $U(1)$ of $E_8$ and a $U(1)$ acting on the extra matter, and it would not increase $r_{\text{flavor}}$. The geometric flavor rank \eqref{flavor rank value} is thus compatible with a decoupled theory.\\
\indent The compatibility of our model with a factorized theory of the form \eqref{decoupled theory} can be sharpened by deforming the threefold \eqref{pagoda tate}, splitting the pagoda singular point on $C$ into $p$ conifold points \textit{off} $C$. This is achieved by modifying $a_3$, and keeping the other parameters identical with respect to \eqref{parameters choice}:
\begin{equation}\label{conifold points}
    a_1 = 0, \quad a_2 = c_1s_2^2,\quad a_3= s_1 s_2^2+\epsilon s_2^3, \quad a_4 = s_1s_2^3+c_2 s_1^4+c_3\tau^p s_2^{p+4}, \quad a_6 = 0.
\end{equation}
Picking $\epsilon = 0$ recovers the pagoda model. With $\epsilon\neq 0$, it is easy to check that the threefold is singular at $p$ isolated conifold points, all lying outside the curve $C$ (namely at non-vanishing values of $\tau$), while $C$ hosts twelve $I_1$ fibers as in the ordinary E-string: this manifestly engineers a decoupled theory of type \eqref{decoupled theory}. The same small resolution as in Section \ref{sec: resolution} resolves each conifold point, producing $p$ exceptional curves $b_1,\dots,b_p$. These are all homologous to the pagoda curve $b$. We can see this by noticing that the resolved threefolds form a smooth family in $\epsilon$, so homology classes do not jump, and as $\epsilon\to0$ the $p$ conifold points collide at $s_1=\tau=0$ and each $b_i$ degenerates to $b$. Since $b$ lies in $D$, the classes $[b_i]=[b]$ do not add new generators to $H_2$, and the flavor rank is $r_{\text{flavor}}=9$ also for $\epsilon\neq0$. This makes manifest the fact that the SCFT and matter sectors (which are decoupled in this example) can be charged under the same flavor $U(1)$ symmetry\footnote{Notice that also the anomaly polynomial cannot distinguish between the scenario in which the extra matter is coupled to the E-string, and the one in which it is not. On the tensor branch, the anomaly polynomial is fixed by topological data of the resolved threefold, which are constant along the smooth family \eqref{conifold points}: for $\epsilon\neq0$  the threefold engineers a manifestly decoupled theory. Explicitly, computing the anomaly polynomial on the tensor branch \cite{Ohmori:2014kda}, we see that the extra matter carries no string charge, since $D\cdot b=K_D\cdot b=0$: the first equality follows from adjunction \eqref{KD adjunction}, and the second from \eqref{K fiber} and \eqref{b fiber}. Hence the extra matter does not modify the Green-Schwarz term, and simply adds its own contribution to that of the ordinary E-string, exactly as in a decoupled theory with a shared $U(1)$ background.}.

\subsection{A cascade of decorated 5d theories from flopping curves}\label{sec: RG cascade}

In this section we flop curves out of the divisor of the 5d KK theory arising from the circle reduction of the decorated E-string, finding a cascade of decorated 5d SCFTs. The 5d KK theory is not a 5d SCFT, since its UV completion is the 6d theory. As for the ordinary E-string, genuine 5d SCFTs are obtained by decoupling hypermultiplets, namely by flopping curves $C_i$ (specified in \eqref{flopping curves} below) with normal bundle $\mathcal O(-1)\oplus\mathcal O(-1)$ out of the compact divisor and sending their volumes to infinity. In this limit the KK modes decouple as well: each flopped curve is a section of $D\to C$, and therefore meets every elliptic fiber once. After the first flop the fiber class splits as 
\begin{equation}
   [F]=[F']+[C_i^+], 
\end{equation}
where $F'$ is the image of $F$ in the new compact divisor and $C_i^+$ is the curve after the flop. Hence $1/R\propto{\rm vol}(F)\geq{\rm vol}(C_i^+)$, so that sending ${\rm vol}(C_i^+)\to\infty$ also sends the KK scale to infinity. Moreover, since $F$ is no longer contained in the compact divisor, the latter can be shrunk without recovering the 6d theory. We would also like the curves undergoing the flop to be orthogonal to $b$ (so that each flop is an isomorphism in the neighborhood of $b$, and the pagoda germ, carrying the extra matter, is unaffected).\\
\indent To identify them, we pick an explicit representative for the exceptional curve $b$. Since the isometries of $\text{Pic}(D)$ fixing $K_D$ act transitively on the roots, we can pick the basis \eqref{basis} such that the representative for the curve $b$ is:
\begin{equation}\label{b choice}
    b = E_8-E_9.
\end{equation}
All the quantities computed below are intersection numbers, and are therefore independent of the choice \eqref{b choice}.
It is easy to check that \eqref{b fiber} and \eqref{b square} are satisfied by this choice. Given that $b$ and $b'$ are components of the same fiber, we have $b+b' =  F$, and:
\begin{equation}
    b' = 3H-E_1-E_2-E_3-E_4-E_5-E_6-E_7-2E_8.
\end{equation}
We then pick the following curves:
\begin{equation}\label{flopping curves}
    C_i = E_i\,\, (i=1,\ldots,7), \quad C_8 = H-E_8-E_9.
\end{equation}
Using the fact that the canonical bundle of the surface is $K_D = -F$, we get:
\begin{equation}
    K_D \cdot C_i = -1,
\end{equation}
which implies that the curves $C_i$ meet the elliptic fiber exactly once, as expected.
Since $C_i^2=-1$ in $D$ and $K_D\cdot C_i = -1$, both directions normal to $C_i$ in $\tilde{X}_P$ have degree $-1$. Consequently $N_{C_i/\tilde{X}_P} =\mathcal{O}(-1)\oplus\mathcal{O}(-1)$, as an extension of $\mathcal{O}(-1)$ by itself on $\mathbb{P}^1$ always splits, and $C_i$ can be flopped. We can then perform the flops sequentially: crucially, they do not affect $b$, which is orthogonal to all the $C_i$. We obtain the renormalization-group cascade in Table \eqref{RG cascade}.
\begin{equation}\label{RG cascade}
    \begin{array}{c|c|c|c}
      \# \textbf{ of flopped curves }j   & \textbf{Surface} & \textbf{Mass parameters} & \textbf{5d theory} \\
      \hline
     0  & dP_9 & 9 & \text{decorated E-string KK theory} \\
     1,\ldots,7 & \text{weak }dP_{9-j} & 9-j & \text{decorated 5d $E_{9-j}$ theory} \\
     8 & \mathbb{F}_2 & 1 & \text{decorated 5d $E_1$ theory}
    \end{array}
\end{equation}
It is interesting to observe that the compact surface for $j=1,\ldots,7$ is a \textit{weak} del Pezzo surface, due to the presence of the curve $b$, which has $K_D\cdot b = 0$, as we have seen in \eqref{b square}. Shrinking it would indeed produce an isolated singularity of type $A_1$ \textit{on the surface} (but only an isolated singularity of pagoda type in the threefold). Therefore the surface is insensitive to the data ``$p$'' that encodes the terminal singularity \eqref{pagoda patch}, whose germ is visible only in the threefold. This is to be contrasted with the usual interpretation of the 5d $E_n$ theories, which can also be engineered via weak del Pezzo surfaces corresponding to the threefold geometries $\text{Tot}(K_D)$: in that case, the $-2$ curve is free to move in the direction orthogonal to the divisor, and hence produces a whole non-compact line of $A_1$ singularities in the threefold.\\
\indent The number of mass parameters of the 5d theories in \eqref{RG cascade} is the same as for the ordinary 5d $E_n$ counterparts, despite the presence of an extra matter sector with a $p$-dimensional Higgs branch, from M2-branes wrapping the curve $b$: by the argument given below \eqref{flavor rank}, this is in principle compatible with the extra matter being decoupled from the 5d interacting sector.
Besides, the divisor data is unchanged with respect to the standard 5d $E_n$ counterparts. In particular, $D\cdot b=K_D\cdot b=0$: the extra matter sector is neutral under the dynamical $U(1)$, its mass ${\rm vol}(b)$ is a pure flavor parameter, and the Coulomb-branch-dependent part of the prepotential does not depend on $p$. This implies that the weak $dP_{9-j}$ surfaces (for $j=1,\ldots, 7$) and $\mathbb F_2$ for $j=8$ can be shrunk to zero size, reaching a superconformal fixed point. \\
\indent It is important to remark how these are related to the ordinary 5d $E_{9-j}$ SCFTs: the distinction lies in the matter content of the class $b$: in the decorated 5d SCFTs there is extra matter localized at the exceptional curve of the terminal pagoda singularity, whereas in the geometries $\text{Tot}(K_D)$ of the ordinary $E_{9-j}$ SCFTs the same class moves along a non-compact line, giving rise to an explicit $A_1$ flavor sector and no extra matter.

The endpoint of the cascade \eqref{RG cascade} warrants a further comment: after flopping $E_1,\ldots,E_7$ out of the compact divisor, one is left with the curve lattice:
\begin{equation}
    \text{Pic}(\text{weak }dP_2) = \langle H,E_8,E_9\rangle.
\end{equation}
The only $(-1,-1)$ curve that is orthogonal to $b$ is then $H-E_8-E_9$. The surviving curves after the flop are:
\begin{equation}
    b=E_8-E_9, \quad u = H-E_8,
\end{equation}
where $b$ and $u$ span the sublattice orthogonal to all the flopped curves. We compute the intersection pairing between $b$ and $u$:
\begin{equation}
    b^2 = -2, \quad u^2 = 0, \quad b\cdot u = 1.
\end{equation}
Since $b$ is an effective $(-2)$-curve, this identifies the surface as $\mathbb F_2$. The $\mathbb F_2$ compact divisor, containing the exceptional curve of a Reid pagoda of order $p$, has the same local data (divisor, normal bundle and pagoda germ along $b$) as the resolution of the $\mathbb Z_2$ orbifold of the Reid pagoda of order $2p+1$ constructed and analyzed in \cite{Collinucci:2025rrh}. Indeed, the only mass parameter left (namely, orthogonal to $K_{\mathbb F_2}$) is the volume of the curve $b$, which is identified with the inverse square of the Yang-Mills coupling of the $SU(2)$ theory, as shown in \cite{Collinucci:2025rrh}. In \cite{Collinucci:2025rrh}, vevs of the extra matter are shown to obstruct this Kähler modulus, freezing the gauge coupling at infinity. In our language this follows from the identification of the extra matter mass with ${\rm vol}(b)$, and would hold equally for decoupled matter charged under the same diagonal $U(1)$, as we have argued in Section \ref{sec: flavor rank}. Moreover, \cite{Collinucci:2025rrh} interpret this theory as local $\mathbb F_2$ deformed by a non-constant $SU(2)$ flavor background, which is the 5d counterpart of the role of $c_3$ discussed right before Section \ref{sec: resolution}. Finally, the endpoint of the cascade is constrained: flops must avoid $b$, and $\mathbb F_2$ is the only weak del Pezzo surface of degree 8 containing a $(-2)$-curve, so no decorated versions of the $\tilde E_1$ or $E_0$ theories arise. This theory is thus the endpoint of a whole family of decorated rank-1 5d theories.

\section{Outlook}\label{sec: outlook}
In this work we have constructed E-string SCFTs decorated by a specific type of terminal singularities admitting a small resolution, known as ``Reid's pagodas''. This shows that the gap observed in 5d \cite{Closset:2020scj,Collinucci:2025rrh} is present in a similar form in 6d: the tensor-branch geometries of 6d SCFTs are not exhausted by the data classified in \cite{Heckman:2013pva,Heckman:2015bfa}. No conceptual obstacle prevents us from realizing analogous decorated versions of the E-string employing different kinds of terminal singularities (for a partial classification of such threefolds, see e.g. \cite{reidyoungpersonsing,Katz:1992aa,Wang:2015mra,DeMarco:2022dgh,Collinucci:2022rii}), which in general can admit  multiple resolved $\mathbb{P}^1$'s (thus generating matter charged under some abelian flavor group with rank greater than one) or no small resolution (in which case the matter sector is uncharged under the flavor group\footnote{The non-resolvable case requires more care, since the extra matter is not charged under any geometric flavor $U(1)$, and the flavor-rank considerations of Section \ref{sec: 5d SCFTs} do not apply.}). More broadly, one could envisage implementing decorations by matter sectors supplied by terminal singularities for 6d SCFTs of arbitrary ranks.
Adopting this view, it would be tempting to advocate that the correct complete classification of 6d SCFTs (and, presumably, of 5d SCFTs alike, employing standard circle reduction arguments) is obtained by enumerating non-trivial interacting field theories, together with decorations of extra matter sectors engineered by terminal singularities. In the context of 5d SCFTs, hints in this direction were already provided in \cite{Closset:2020scj} and more recently in \cite{Collinucci:2025rrh}. We leave the exploration of the 6d perspective presented here for future systematic work. This would of course only make sense for the classification of SCFTs if these extra matter sectors are coupled to the SCFT.

\section*{Acknowledgments}
The author would like to thank Roberto Valandro and Michele Del Zotto for insightful comments on an earlier version of the draft, as well as Timo Weigand for valuable discussions.
The research of the author is funded by the VR Centre for Geometry and Physics (VR grant No.\ 2022-06593) at Uppsala University.

\bibliographystyle{at}
\bibliography{bibliography.bib}

\end{document}